\documentclass[11pt]{article}
\usepackage[normalem]{ulem}
\pdfoutput=1 

\usepackage{jheppub} 
\usepackage{url}
\usepackage{caption}
\usepackage{subcaption}
\usepackage{extarrows}

\usepackage[latin9]{inputenc}
\usepackage{float}
\usepackage{amsmath}
\usepackage{amssymb}
\usepackage{graphicx}
\usepackage{esint}
\usepackage{hyperref}
\usepackage{comment}
\usepackage{color}
\usepackage{microtype}
\usepackage{cleveref}
\usepackage{breakurl}
\usepackage{bbm}
\newcommand{\be}{\begin{equation}}
\newcommand{\ee}{\end{equation}}
\newcommand{\ben}{\begin{displaymath}}
\newcommand{\een}{\end{displaymath}}
\newcommand{\bea}{\begin{eqnarray}}
\newcommand{\eea}{\end{eqnarray}}
\def\K{K{\"a}hler }
   \newcommand{\rf}[1]{(\ref{#1})}
\newcommand{\vp}{\varphi}

\def\be{\begin{equation}}
\def\ee{\end{equation}}
\def\bea{\begin{eqnarray}}
\def\eea{\end{eqnarray}}
\def\ba{\begin{array}}
\def\ea{\end{array}}
\def\bit{\begin{itemize}}
\def\eit{\end{itemize}}

\def\a{\alpha}

\def\vp{\varphi}

 \makeatletter

\allowdisplaybreaks

\makeatother

\makeatletter
\DeclareRobustCommand{\rcite}[1]{%
  \rcite@aux#1,\@nil{#1}%
}
\def\rcite@aux#1,#2\@nil#3{%
  \if\relax#2\relax
    Ref.~\cite{#3}%
  \else
    Refs.~\cite{#3}%
  \fi
}
\makeatother

\hypersetup{
    colorlinks = true,
    citecolor = {blue},
    linkcolor = {blue},
    urlcolor = {blue},
}

\def\be{\begin{equation}}
\def\ee{\end{equation}}
\def\eqn#1{eq.~\eqref{#1}}

\def\rcite#1{ref.~\cite{#1}}

 \title{\LARGE{ \boldmath{New Exponential and Polynomial $\xi$-attractors}}}

\author{Renata Kallosh and Andrei Linde}

\affiliation{Leinweber Institute for Theoretical Physics at Stanford, 382 Via Pueblo, Stanford, CA 94305, USA}
\emailAdd{kallosh@stanford.edu}
 \emailAdd{alinde@stanford.edu}

\abstract{   We introduce a new family of cosmological attractors with non-minimal coupling of gravity and non-canonical kinetic terms. In the Einstein frame, these models transform into a class of exponential and polynomial attractors with the spectral index $n_{s}$ spanning a broad range $1-2/N \leq n_{s} < 1-1/N$, and $r$ can decrease to zero in the limit  $\xi \to \infty$.  This is sufficient to match any combination of Planck, BICEP/Keck, ACT, SPT, and DESI  data. We present a supergravity implementation of these models.
 }

\begin{document}

\maketitle

 \parskip 5pt

\section{Introduction}

Cosmological attractors generalize the Starobinsky and Higgs inflation models, while providing stability of cosmological predictions with respect to modifications of the inflaton potential. The first models of that type were the $\xi$-attractor models, based on models with a non-minimal coupling of the scalar field to gravity \cite{Kallosh:2013tua}.  In the strong coupling (large $\xi$) limit, these models predict $n_{s} = 1-2/N$ and $r = 12/N^{2}$, just as the Starobinsky and Higgs inflation models.

A more general class of cosmological attractor models is the class of $\alpha$-attractors, which is motivated by supergravity in hyperbolic moduli space \cite{Ferrara:2013rsa,Kallosh:2013yoa}. These models (T-models and E-models) predict the same value of $n_{s} = 1-2/N$, but allow for a large set of values of $r = 12\alpha/N^{2}$. In all of  these models, inflationary potential approaches a plateau exponentially, 
\be
V_{\rm exp}(\vp)= V_0( 1-e^{-{\sqrt{2\over 3\a} \vp}}+\cdots)\, ,
\label{exp}\ee
so one can call these models exponential attractors with observables 
 at large $N$
\be
n_{s} = 1-{2\over N} \ , \qquad  r = {12 \a\over  N^{2}}
\label{a}\ee
 Such potentials can also be found in a special class of models with non-minimal coupling to gravity \cite{Galante:2014ifa}.  
 
These models provided a good match to CMB-related observational data. The most recent analysis of the Planck, ACT, SPT, and BICEP/Keck \cite{Balkenhol:2025wms} suggests that 
\be
n_s= 0.9682\pm 0.0032\, ,\qquad r<0.034 \ .
\label{Planck}\ee
These results are consistent with the exponential attractors discussed above.

However, the situation changed after ACT \cite{AtacamaCosmologyTelescope:2025blo} and SPT \cite{SPT-3G:2025bzu} combined all available CMB data with that of DESI \cite{DESI:2025zgx}. This favored a greater value of $n_s$  \cite{Balkenhol:2025wms}:
\be
n_s= 0.9728\pm 0.0029\, ,\qquad r<0.035 \ .
\label{andD}\ee
It was noticed in \cite{SPT-3G:2025bzu,Ferreira:2025lrd,McDonough:2025lzo} and \cite{Balkenhol:2025wms} that combining CMB and DESI results to constrain $n_{s}$ is potentially problematic, as these datasets are in $2\sigma$ to $4\sigma$  tension with each other.   Until this issue is resolved, it is reasonable to continue studying benchmark models such as the Starobinsky model, Higgs inflation, and exponential $\xi$ and $\alpha$-attractors, while simultaneously exploring other inflationary models
matching the $n_{s}$ constraints \rf{andD}.

The first model addressing this issue was based on the theory of $\xi$-attractors with the quadratic Jordan frame potential with $\xi = O(1) $ \cite{Kallosh:2025rni}: 
 \begin{equation} 
{1\over \sqrt{-g}} \mathcal{L} =   \frac12 (1+ \xi \phi) R  - \frac12 (\partial \phi)^2 -  \frac12  m^{2}\phi^{2}  \  . \label{simple} 
 \end{equation}
This model in the large $N $ limit and  $\xi = O(1) $ predicts 
\begin{equation}\label{nsrxi1}
n_s \approx 1- {3\over 2 N},   \qquad  r\approx {4\over \xi \, N^{3/2}}\ .
\end{equation}
This is a much better match to the CMB-DESI data. The potential of this model in the Einstein frame at large values of the canonically normalized inflaton $\vp$  at $\xi=1$ is given by
 \be
V = {m^{2}\over 2} \left(1 -{8\vp^{-2}}+O(\vp^{-4})\right) \ .
\label{largephi}\ee
 Thus, the potential approaches the plateau polynomially.  

Ref. \cite{Kallosh:2025rni} was followed by a large set of other proposals to match the CMB-DESI data, see a review of these developments in \cite{Kallosh:2025ijd}. The general conclusion is that one can match the CMB-DESI data  in the theory of polynomial attractors (P models) with potentials for the canonical field $\vp \to \infty$ of the form
\be
V = V_0\, \Big(1- \Big ({\mu\over  \vp }\Big)^{k}+\dots \Big ) \ .
\label{polynomial}\ee
These models predict  at small $\mu$ 
\be\label{KKLTIw}\
  n_{s} = 1-{2(k+1)\over (k+2)N} \, , ~~r = 8 \, k^2 \mu^{2k\over k+2} \Big({1\over  k  (k+2) N)}\Big)^{\frac{2 (k+1)}{k+2}}   
\ee  
which spans the broad range of $n_{s}$ from $1-{2\over N}$ to $1-{1\over N}$ \cite{Martin:2013tda, Kallosh:2018zsi,Kallosh:2019hzo,Kallosh:2022feu}.

In this paper, we will construct a new class of inflationary $\xi$-attractors with non-minimal coupling of scalars to gravity.  We will show that in this class of models it is possible to reproduce the results of all previously known single-field exponential and polynomial attractors,  
including E  and T-models of exponential $\alpha$-attractors \cite{Kallosh:2013yoa}, and P models: polynomial $\a$ attractors  and KKLTI-type attractors \cite{Kallosh:2018zsi,Kallosh:2019hzo,Kallosh:2022feu,Martin:2013tda}. We present supergravity implementations of these models, following \cite{Kallosh:2026tjm}.

\section{General setting}\label{Sec:Set}

Consider a general class of  models with non-minimal coupling to gravity  in the Jordan frame \cite{Galante:2014ifa}: 
 \begin{align}\label{Jordan}
{\mathcal{L}_{\rm J}\over \sqrt{-g_J}} = \  \frac12  \Omega( \phi) R  - \frac12 K_J(\phi ) (\partial \phi )^2 -  V_J(\phi ) \ .
\end{align}
We perform a Weyl transformation of the metric 
\be
g_{\mu\nu}^J \to \Omega \, g_{\mu\nu} \ .
\label{m}\ee
The Lagrangian in the Einstein frame becomes
 \begin{align}
 {\mathcal{L}_{\rm E}\over \sqrt{-g}} =  \frac12   R - \frac12 K_E(\phi ) (\partial \phi )^2    - V_E(\phi )    \ ,  
 \label{Einstein}
 \end{align}
 where
\be
K_E(\phi )={K_J(\phi ) \over \Omega(\phi )} + {3\over 2} {(\Omega')^2\over \Omega^{2}}, \label{KE}
\ee
and
\be
\quad 
   V_E(\phi ) =  \frac{V_J(\phi )}{\Omega^2(\phi )}  \ .
\ee
In many cosmological models describing scalar fields with nonminimal coupling to gravity, such as Higgs inflation \cite{Salopek:1988qh,Bezrukov:2007ep} and $\xi$-attractors \cite{Kallosh:2013tua}, the authors adopt the simplest choice $K_J(\phi) = 1$. The choice of $K_J=1$ for Higgs inflation reflects the fact that Higgs scalars in the absence of gravity have a flat geometry. One of the consequences of this choice is that in the large $\xi$ limit the first term in \rf{KE} disappears, and therefore $K_E(\phi )$ reduces to ${3\over 2} {(\Omega')^2\over \Omega^{2}}$. That is the reason why $\xi$-attractors, as well as the Higgs inflation models, have the universal predictions $n_{s} = 1-2/N$ and $r = 12/N^{2}$ in the large $\xi$ limit.

However, sometimes a more judicious choice of the function $K_J(\phi )$ can lead to new interesting possibilities. 
For example, the choice $K_J(\phi )= {1\over 4\xi} {(\Omega')^2\over \Omega}$ makes both of the terms in the expression  $K_E(\phi )={K_J(\phi ) \over \Omega(\phi )} + \frac32 (\ln \Omega(\phi ))'^2$ proportional to each other. This allowed us to implement $\alpha$-attractors in the context of theories with non-minimal coupling to gravity  \cite{Galante:2014ifa}.

In this paper, we will make a similarly simple but important choice of the Jordan frame functions $K_J(\phi )$, $V_J(\phi)$. We will allow the choice of these functions to be determined by our choices of the Einstein frame functions $K_E(\phi )$  and $V_E(\phi)$ \rf{KE}. In order to do it, we will take 
\be
K_J(\phi) = K_E\, \Omega -{3\over 2} {(\Omega')^2\over \Omega}\, , \quad 
 V_J=    V_E \, \Omega^2\  . 
\label{newKVJ} \ee
 We will look for potentials $V_{E}$ approaching a plateau at large values of the canonically normalized field $\vp$. 
 
This choice may seem tautological because equations \rf{KE} and \rf{newKVJ} are equivalent. 
However, in this paper, we will focus on inflationary models with non-minimal coupling to gravity that can be implemented in supergravity. The only known way to do it is to use the superconformal formulation of supergravity, in which the Jordan frame and the Einstein frame appear as two different gauges of the same theory \cite{Kallosh:2000ve,Ferrara:2010yw,Ferrara:2010in}.  

In the superconformal approach, one begins with two independent functions of complex scalar fields: the frame function $\Omega(z, \bar z)$ and the \K potential $K(z, \bar z)$. The function $K_{E}(\phi)$ in \rf{Einstein} is the second derivative of the  \K  potential at $z=\bar z=\phi$.  If one knows the  \K  potential, $K(z, \bar z)$, and $\Omega(z, \bar z)$, one can always find $K_{J}(z, \bar z)$  \cite{Kallosh:2026tjm}. At   $z=\bar z=\phi$ it is, in fact, given in eq. \rf{newKVJ}.

In  \cite{Kallosh:2026tjm} we have replaced the generic textbook moduli coordinate $(z, \bar z)$ in \cite{Freedman:2012zz} by $(T, \overline{T})$ since in hyperbolic geometry $\a$ attractors (besides the disk variables $Z\bar Z<1$ \cite{Kallosh:2013yoa}) one can use  also  a  half-plane modulus $T$ with $T+\overline{T}>0$ \cite{Cecotti:2014ipa,Kallosh:2015zsa} which allows a simple description of all exponential and polynomial $\xi$ attractors using
\be
 K(T, \overline{T})= -{1\over 2 \xi} \ln (T+\overline{T})  \ .
\ee
In general polynomial attractors $(\mu, k)$ in supergravity in \cite{Kallosh:2025dac} we have have also used a $(T, \overline{T})$ variable with   
\be
K(T, \overline{T}) =   {2\over \xi (2-n)^{2}}  ( {T} \overline{T} )^{2-n\over 2} \ .
  \label{KkkK} \ee
That is why,  in this paper, with  $T=z=\bar z$,
we will specify inflationary models by choice of $\Omega(T)$, $K_{E}(T)$, and $V_{E}(T)$. Then, for each of these models, we will present its supergravity version following the lines of \cite{Kallosh:2026tjm}.
The Jordan frame Lagrangian  ${\mathcal{L}_{\rm J}\over \sqrt{-g_J}}$ in eq. \rf{Jordan} for any choice of the frame function  $\Omega(T)$   is 
\be\boxed {
{\mathcal{L}_{J}\over \sqrt{-g}_J} = \frac12  \Omega \, R  - \frac12 \left (K_E \, \Omega  \, -{3\over 2} {(\Omega')^2\over \Omega}\right ) (\partial T )^2 -  \Omega^2V_E(T )  \ .}
\label{JordanT} \ee
Cosmological models that we will describe will be given in the form  \eqn{JordanT} either in the Jordan gauge $\Omega^J(T)$ or in the Einstein gauge $\Omega^E(T) = 1$.

In the Einstein frame our action \rf{JordanT} at $ \Omega^E(T)=1$ is
\be
{\mathcal{L}_{E}\over \sqrt{-g}_E} = \frac12   R  - \frac12  K_E  (\partial T )^2 -  V_E(T )  \ .
\label{Ein} \ee
This action is a different superconformal gauge from the one in eq. \rf{JordanT}. One may use, for example, $\Omega^J(T)= 1+ \xi T ^{n}$ in \rf{JordanT}, which is a convenient choice of non-linear scalar-gravity coupling in cosmological models.  However, the Einstein frame action  \rf{Ein} does not depend on the choice of $\Omega(T)$ in the Jordan frame action \rf{JordanT}; it depends only on the choice of $K_E$ and $V_{E}$. 

In this paper, we will use $K_{E} \sim T^{-n}$. We will show that with this choice, and with a proper choice of the potential $V_{E}$, the theory \rf{JordanT},  \rf{Ein} will describe a broad family of pole inflation cosmological attractors \cite{Galante:2014ifa,Terada:2016nqg,Kallosh:2019hzo}. 

Our investigation of the supergravity implementation of such models will follow the recent paper \cite{Kallosh:2026tjm}, as well as the paper \cite{Kallosh:2025dac}, which explains how one can construct supergravity models with an arbitrary \K potential $K(T,\overline{T})$ and an arbitrary potential $V(T)$.

\section{\bf{\boldmath New $(\xi, n )$-attractors}}\label{Sec:newxi}

\subsection{\boldmath Hyperbolic geometry models, $n=2, \xi={1\over 6\a}$ } \label{Sec:A} 

Here we will discuss the models with $n = 2$, corresponding to hyperbolic geometry of the moduli space.\footnote{This choice of $K_E$ corresponds to a supergravity version of this class of models in the Einstein frame, see eq.  \rf{hyperE} below at $T=\overline{T}$.}  
\be
 K_E (T)={1\over 4\, \xi \, T^2} \,  , \quad V_J=    V_E \, \Omega^2 \ .
\label{choice}\ee

The Jordan frame action  \rf{JordanT} is 
\be\label{JordanTa}
{\mathcal{L}_{J}\over \sqrt{-g}_J} = \frac12  \Omega \, R  - \frac12 \left ( {\Omega\over 4\, \xi \, T^2}  \, -{3\over 2} {(\Omega')^2\over \Omega}\right ) (\partial T )^2 -  \Omega^2V_E(T ) \ .
\ee
In particular, for $\Omega(T)=1+\xi T^2 $ we have
\bea
\label{JordanTa2}
 {\mathcal{L}_{J}\over \sqrt{-g}_J} =   \frac12  (1+ \xi \, T ^2) R  - \frac12 \left ({1+ \xi \, T ^2\over 4\, \xi \, T^2}  -  {(\xi T)^2\over 1+ \xi \, T ^2}\right ) (\partial T )^2 
-  (1+\xi T^2)^2 V_E(T )  \ .
\eea
However, the Einstein frame action corresponding to the Jordan frame action \rf{JordanTa} does not depend on the choice of $\Omega(T)$ in \rf{JordanTa}, \rf{JordanTa2}:
\be
{\mathcal{L}_{E}\over \sqrt{-g}_E}  =    \frac{1}{2} {R}  - {1\over 8\, \xi}  {( {\partial} T)^2 \over  \, T^2}  - V_E (T) \ .
  \label{Einstxi}
\ee
Therefore, the observational consequences of the broad range of the models \rf{JordanTa}  depend only on the choice of the functions $K_{E}$ and $V_{E}$.

  For the simplest exponential $\xi$-attractor E-model, T-model,  are given by the following expressions\footnote{These potentials correspond to the choices for these models in supergravity made in \cite{Kallosh:2026tjm}. }
  \be
V_E^{E-model}(T) = V_{0} \,   (1-T) ^{2m} \ , \qquad 
 V_E^{T-model} (T)=V_{0} \, \left ( {1-T\over 1+T} \right )^{2m} \ ,
\label{T}\ee
and for the polynomial $\xi$-attractor models
\be
 V^{polynomial}_{E}(T) =V_{0} \, { 1\over  1+ (\ln^2 T)^{-m}}  \ .
\label{pol}
\ee

One can switch in eq. \rf{Einstxi} to a canonical variable $\varphi $:
\be
T=\overline{T}=e^{-{2\sqrt {\xi}  } \varphi } \ ,
\label{alpha}\ee
which brings the action in eq. \rf{Einstxi} to the form
\be
{\mathcal{L}_{E}\over \sqrt{-g}_E} (\vp) =    \frac{1}{2} {R}  -  {1\over 2} ( {\partial} \vp)^2- V_E (\vp) \ .
\ee
The potentials for the new exponential  $\xi$-attractors (E-models and T-models) as  functions of the canonical field $\vp$ are
\be
V_E^{E\, model}(\vp) =  V_0 \big (1-e^{-2{\sqrt \xi} \varphi }  \big)^{2m}   \ ,
\qquad 
V_E^{T\, model}(\vp) =  V_0  \tanh^{2m} {{\sqrt \xi} \varphi }   \ ,
\label{hyperTcan}\ee 

The Einstein frame action of $\alpha$-attractors is known to be 
\begin{eqnarray}
{\mathcal{L}_{E}\over \sqrt{-g}_E} (T) =    \frac{1}{2} {R}  - {3\a\over 4} {(  {\partial} T)^2 
\over  T^2}  - V_E (T) \ .
  \label{al}
\end{eqnarray}
The Einstein frame action of new $\xi$-attractors in eq. \rf{Einstxi} coincides with  the $\a$ attractor action  \rf{al} for 
\be
\alpha={1\over 6\xi} \ .
\ee
One can see the same in the expressions for the potentials; for example, in Eq. \rf{hyperTcan} we have $\sqrt \xi$, replacing the $\sqrt{1\over 6\a}$ in the familiar  $\alpha$-attractor expression $V= V_0 \, \tanh^{2m} {\vp\over \sqrt{6\alpha}} $ \cite{Kallosh:2013yoa}. In the polynomial case \eqn{hyperPcan2} we have $k=2m$.

The potential for the polynomial $\xi$-attractor models \rf{pol} as a function of the canonical field $\vp$ is
\be
V_E^{\rm polynomial} (\vp) = V_0 \  { 1\over  1+  ( 4   \xi \, \vp^2 )^{-m}} \   .
\label{hyperPcan2}\ee 

The corresponding supergravity Einstein action for $\xi$-attractors  is \cite{Kallosh:2026tjm}
\be
{ {\cal L}_E (T, \overline{T})\over \sqrt{-g}} =  {R\over 2} - {1\over 2\,  \xi } \, {\partial T \partial \overline{T}\over (T+\overline{T})^2}-  V_E(T, \overline{T}) \ .
\label{hyperE}\ee 
with \be
V_E^{E\, model}(T, \overline{T}) =V_{0} \, [ (1-T) (1-\overline{T})]^{m} \ , \qquad 
V_E^{T\, model} (T, \overline{T})=V_{0} \, \left [ {1-T\over 1+T} \, { 1-\overline{T}\over 1+\overline{T}} \right ]^{m} \ ,
\label{VET}\ee

\be
V^{polynomial}_E(T, \overline{T}) =V_{0} \, { 1\over  1+ \left(\ln^{2} {T+\overline{T}\over 2}\right)^{-m}}\, \label{VpolA} \ . 
\ee
In the Jordan frame with 
\be
\Omega(T, \overline{T} )= 1+ \xi \, (T \overline{T})^{p/2}
\ee
 it  leads to the Jordan frame action for $\xi$-attractors, which at $T=\overline{T}$ is 
\be
\Omega(T)= 1+ \xi \, T ^p \ .
\ee

\begin{figure}[H]
\centering \includegraphics[scale=0.5]{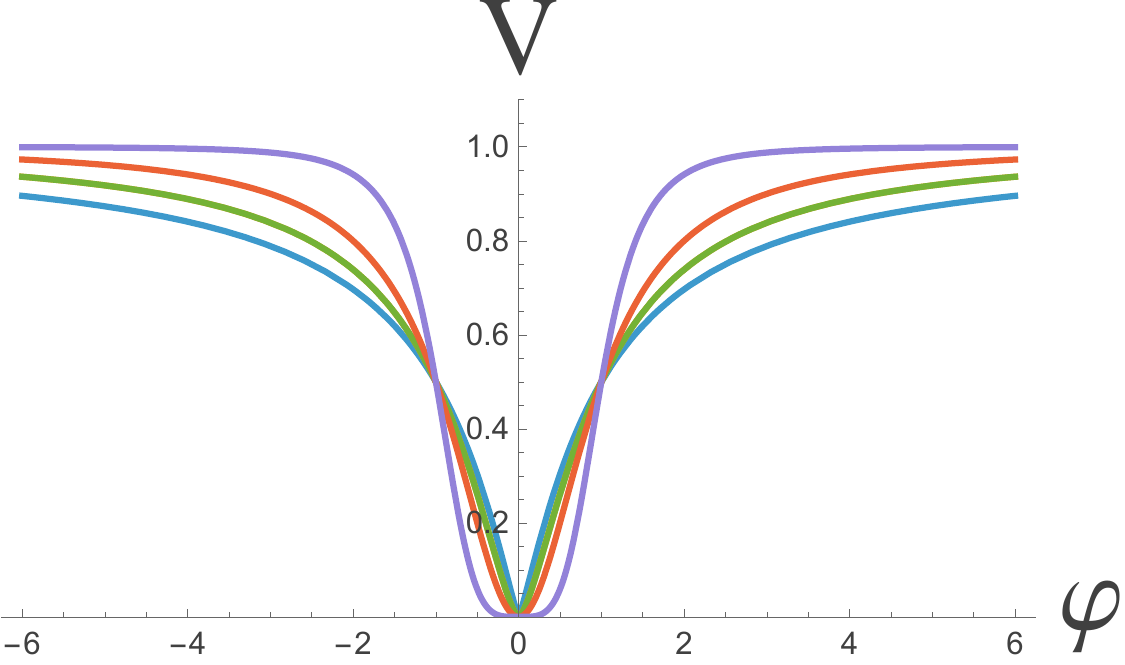}
\vskip -5pt
\caption{\footnotesize  The KKLTI potential \rf{KKLTIk} for $V_{0}=1$ and $\mu = {1\over 2  \sqrt \xi }=1$. The purple, red, green, and blue lines correspond to $ k=4$, 2, 1.5  and  1.2, respectively.  }
\label{KKLTIpot}
\end{figure}

 The potential \rf{hyperPcan2} can also be represented as the  KKLTI polynomial attractor  potential \cite{Kallosh:2018zsi,Kallosh:2019hzo,Kallosh:2022feu,Martin:2013tda}: 
\be\label{KKLTIk}
V_{}(\phi) = V_{0 } {|\vp|^{k}\over |\vp|^{k}+ \mu^{k}}\, , \qquad k = {2m} \, , \qquad  \mu =  {1\over 2  \sqrt \xi } =\sqrt{3\a\over 2} \ .
\ee
We show this potential in Fig. \ref{KKLTIpot} for $\mu = 1$ and several different values of $k$.

\begin{figure}[H]
\centering \includegraphics[scale=0.44]{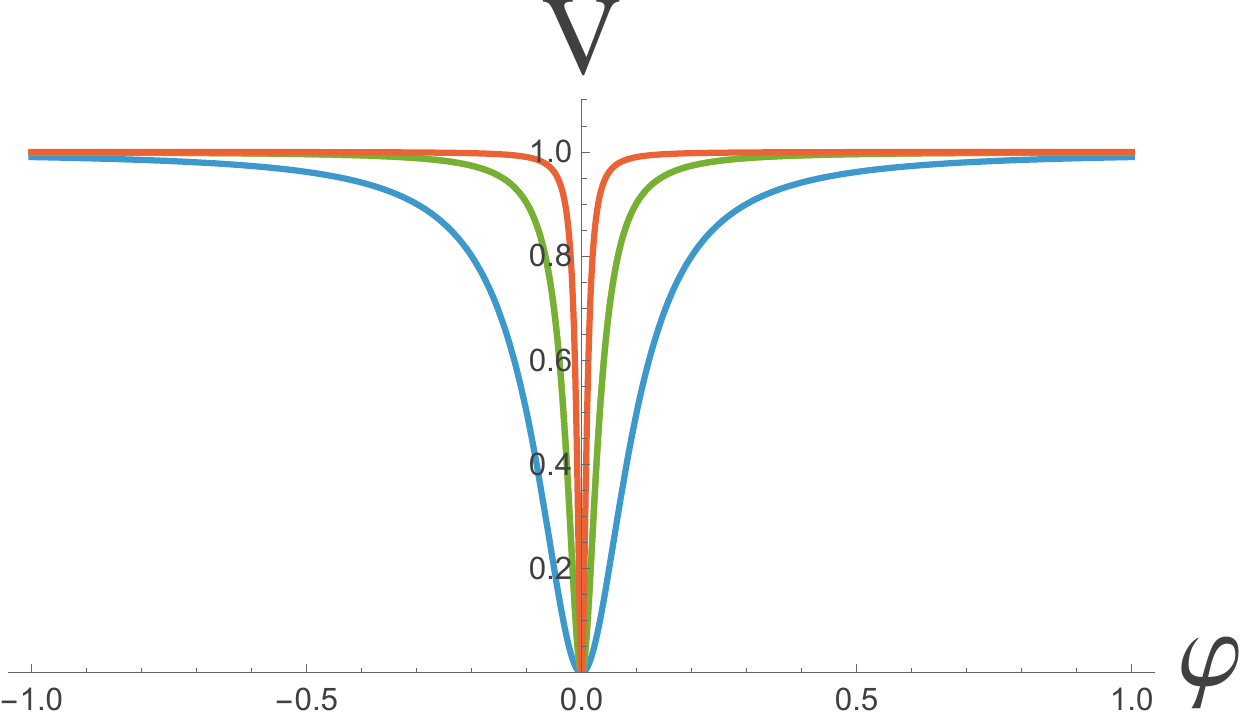}
\vskip -5pt
\caption{\footnotesize  The KKLTI potential \rf{KKLTIk} for $V_{0}=1$, $k = 2$, and various values of $\mu = {1\over 2  \sqrt \xi }$. The blue, green, and red lines correspond to  $\mu = 0.1$,  $\mu = 0.033$,   and $\mu = 0.01$.  }
\label{VarMU}
\end{figure}

It is also instructive to plot these potentials with various values of $\mu = {1\over 2  \sqrt \xi }$, see Fig.~\ref{VarMU}. With an increase of $\xi$ (decrease of $\mu$), the range of $\vp$ corresponding to the last 60 e-foldings of inflation becomes smaller. Therefore, in the large $\xi$ limit, the last 60 e-folds of inflation in these models are described by a small field inflation with $\vp < 1$.

The polynomial $\xi$-attractor potentials \rf{pol}, \rf{hyperPcan2}, \rf{KKLTIk} are well-defined for all  values of $k > 1$, with the attractor (large $\xi$) predictions $n_{s} = 1-{2(k+1)\over (k+2)N}$ \rf{KKLTIw}. This allows to increase $n_{s}$  from the exponential $\alpha$-attractor value $1-2/N$ to $1-4/3N$ (for $k = 1$).  

\begin{figure}[H]
\vspace{-1mm}
\hspace{-3mm}
\begin{center}
\includegraphics[scale=0.33]{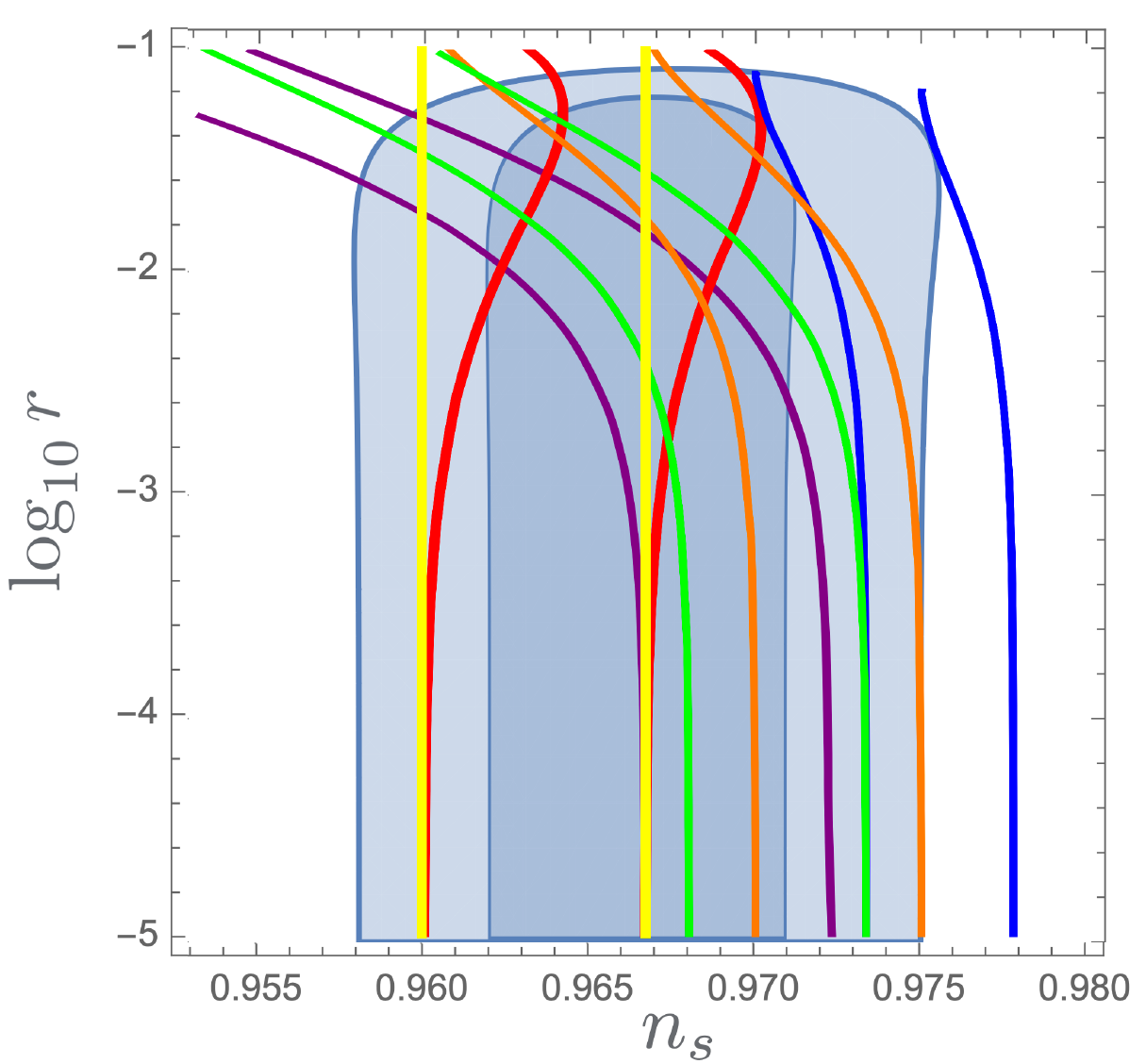} 
\end{center}
\vspace{-.12cm}
\caption{\footnotesize  Predictions of exponential and polynomial $\alpha$-attractors \cite{Kallosh:2019hzo}.  These are equivalent to $n=2$ $\xi$-attractors, exponential and polynomial, with $\xi={1\over 6\a}$ and $\mu^2= {3\a\over 2}$.  Red lines are  $m= 1$ E-models,  yellow lines are $m = 1$ T-models   \rf{hyperTcan}. Polynomial $\xi$-attractors \rf{KKLTIk} with  $k=4,3,2,1$  are shown by yellow, red, purple, green, orange, and blue lines, respectively.   The light blue area represents the Planck 2018 results, incorporating CMB and BAO data.   }
\label{Blue}
\end{figure}

Cosmological predictions of these models for various values of $k$ have been discussed in our paper \cite{Kallosh:2019hzo}; see Fig. \ref{Blue}. This figure illustrates predictions for  $N = 50$ and $60$. The right blue line corresponding to $N = 60$ reaches the attractor value $n_{s} \sim 0.978$. The latest reheating-related constraints prefer slightly smaller $N$, see e.g.   \cite{Martin:2013tda, Drees:2025ngb,Ellis:2025zrf,Iacconi:2025odq,Marciniak:2026unt}.  In particular, for $N = 55$, the family of $\alpha$-attractors and polynomial attractors with $k > 1$ shown in Fig. \ref{Blue} can describe $n_{s}$ in the range $0.964 < n_{s }< 0.976$. For $N= 50$, the related range is $0.96 < n_{s }< 0.973$. These predictions are compatible with the ACT-SPT-DESI constraints \rf{andD}.

At $k \leq 1$, the derivatives of the potential \rf{KKLTIk} diverge at $\vp = 0$, but one can avoid this issue by slightly modifying the potential \rf{VpolA} to obtain the potentials introduced in \cite{Kallosh:2022feu}:  
\be
V^{polynomial}_E(t,\overline{T}) =V_{0} \, { \left(  \ln^{2} {T+\overline{T}\over 2} + 1\right)^{-m}-1\over  \left(  \ln^{2} {T+\overline{T}\over 2} + 1\right)^{-m} +1} \,  \rightarrow \, V_{0}\, {(\vp^{2}+\mu^{2})^{k/2}  - \mu^{k} \over (\vp^{2}+\mu^{2})^{k/2}  + \mu^{k} }\, \label{kk} \ . 
\ee
This potential has a regular, smooth minimum at $\vp = 0$, and the same large $\vp$ behavior as the potential \rf{KKLTIk} for all positive values of $k$.

 Alternatively, one can use potentials
  \be
V^{k}_E(T, \overline{T}) =V_{0} \,    \left(1- {1\over 2}\Big(\ln  {T+\overline{T}\over 2} \Big)^{-k}\right)^{2}\, \label{VpolNew2} \ . 
\ee
This leads to 
\be
V^k(\vp) = V_{0} \left(1-  {1\over 2} \Big(    {\mu \over \vp}\Big  )^k\right )^{2} \ ,
\label{kkk}\ee
where $k = 2m$. 

This class of potentials was introduced in the discussion of pole inflation in \cite{Kallosh:2019hzo}.
These potentials, plotted in Fig. \ref{VarMU},  have a minimum at $\vp= 2^{-1/k} \mu =2^{-1/k}  {1\over 2\sqrt \xi}$, and an infinitely high barrier at $\vp = 0$.  In the large $\vp$ limit, the potential is given by
\be
V^k(\vp) = V_{0} \left(1-   {\mu^{k}\over \vp^{k}}+ ... \right)\ .
\label{kkkl}\ee
 
 \begin{figure}[H]
\centering \includegraphics[scale=0.4]{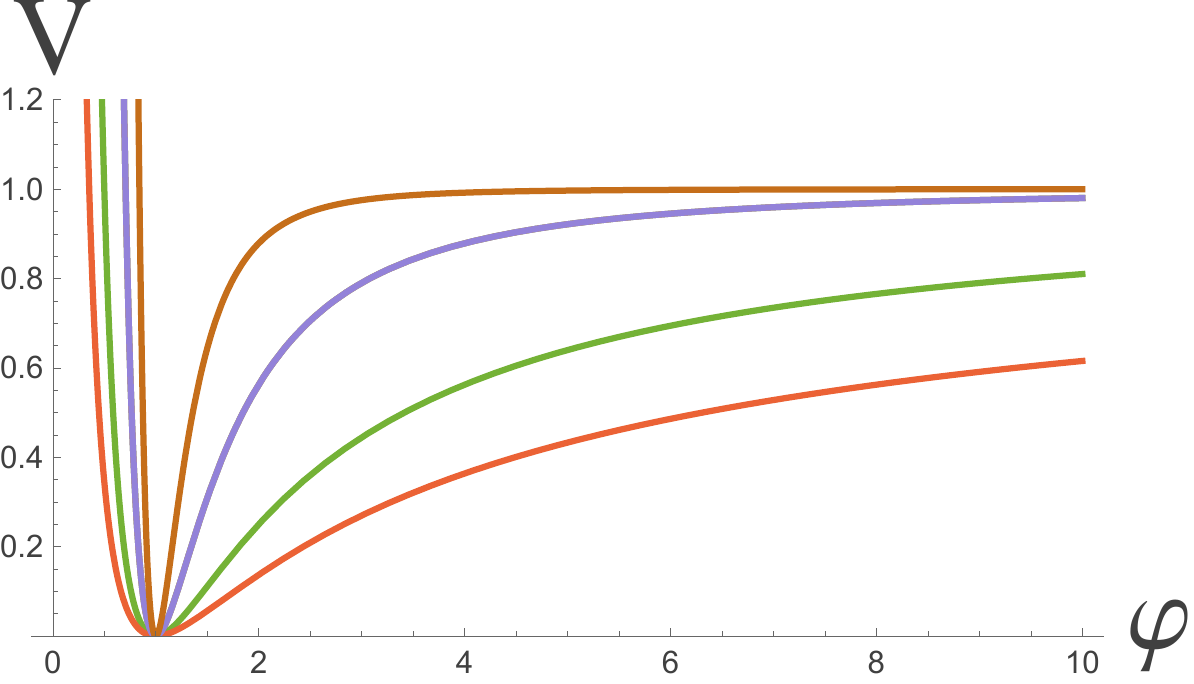}
\vskip -5pt
\caption{\footnotesize  The  potential \rf{kkk} for $V_{0}=1$, $\mu = 2^{1/k}$, and various values of $k$. The brown,  blue, green, and red lines correspond to  $k = 4$,  $2$, $1$, and $2/3$.}
\label{VarMU2}
\end{figure}

The potentials \rf{kk} and \rf{kkk} are well-defined for all  $k>0$. In this class of models, $n_{s}$ can change, as a function of $k$, from $n_{s} = 1-2/N$ (in the large-$k$ limit) to $n_{s} = 1-1/N$ (in the small-$k$ limit). 
In particular, for $N =55$, the values of $n_{s}$ in the attractor regime with large $\xi$ (small $\mu$) span the broad range of values from $n_{s} = 0.964$ to $n_{s} = 0.982$.  
This is more than sufficient to fully cover the ACT-SPT-DESI range $n_s= 0.9728\pm 0.0029$ \rf{andD}.

\subsection{\boldmath General polynomial $(\xi, n \not = 2)$ attractors and KKLTI  models }\label{Sec:B}

Here we will define a new class of $\xi$-attractors with KKLTI potentials  \cite{Kallosh:2018zsi,Kallosh:2019hzo,Kallosh:2022feu,Martin:2013tda}  by making choices of the functions\footnote{This choice corresponds to a supergravity version of this class of models in the Einstein frame.}  
\be
 K_E (T)={1\over  \, \xi \, T^{n}}, \qquad V_{E} = {1\over 1+T} \ ,
\label{choiceA}\ee
we have
\be
  K_J(T) =  { \Omega \over  \, \xi \, T^{n}}\,  -{3\over 2} {(\Omega')^2\over \Omega} \, , \quad    V_J =  \Omega^2 V_E \ .
  \ee
The  Jordan frame action is
\be\label{JordanGeneral}
{\mathcal{L}_{\rm J}\over \sqrt{-g_J}} =  \frac12  \Omega R  -  {1\over 2}\Big ({ \Omega \over  \, \xi \, T^{n}}\,  -{3\over 2} {(\Omega')^2\over \Omega}\Big) (\partial T )^2-  V_0{ \Omega^2\over 1+T  } \,  \ .
\ee
For example, in a model with $ \Omega(T)= 1+\xi T^p$ and $V_{E} = {1\over 1+ T}$  the  Jordan frame action is
\be\label{JordanPolcan}
{\mathcal{L}_{\rm J}\over \sqrt{-g_J}} =  \frac12  (1+\xi T^p) R  -  {1\over 2}\Big (   {1+\xi T^p\over  \xi  {T}^{n}}
 -{3\over 2} {p^2 \xi^2 T^{2(p-1)}\over 1+\xi T^p}\Big) (\partial T )^2-  V_0{(1+\xi T^p) ^2\over 1+T  } \,  \ .
\ee
If we choose a superconformal gauge in the form with $ \Omega(T)= 1+\xi T^n$, we have a Jordan frame action in \eqn{JordanPolcan} with  $p=n$, which depends on two parameters $n, \xi$. This model represents a $(\xi, n)$ attractor.  However, the Einstein frame Lagrangian is given by the general expression   \rf{JordanT} in the Einstein gauge $\Omega^E(T)= 1 $ for any choice of $\Omega(T)$ in the Jordan frame:
\be
{\mathcal{L}_{\rm E}\over \sqrt{-g}} =  \frac12   R  - {  1\over 2 \xi}   {(\partial T)^2 \over  {T} ^{n}}  -   { V_0\over 1+T }\ .
\label{canG}\ee 
Therefore, all observational predictions depend only on the choice of $K_{E}$ and $V_{E}$ \rf{choiceA}. 
The Einstein frame action \rf{canG} is related to the pole inflation attractors \cite{Galante:2014ifa,Terada:2016nqg,Kallosh:2019hzo}, which have been implemented in supergravity in \cite{Kallosh:2025dac}.  

We will begin with the case $n > 2$. In this case, the field $T$ and the canonically normalized inflaton field $\vp$ are related as follows:
\be
T= \left ({n-2\over 2} \sqrt \xi \, \vp\right)^{2\over 2-n} = \left ({ \mu^2\over \vp^2 }\right)^{k\over 2} \ ,
 \ee
where 
\be
k={2\over n-2} \quad \to \quad n= {2(k+1)\over k}>2  \, ,\qquad \mu^2 = {4\over (n-2)^2\xi} \ .
\ee
In this case, when $n$ changes in the range $2<n<\infty$, the parameter $k$ scans the full range $0< k < \infty$.  

Our canonical field $\vp$ has the potential 
\be\label{KKLTcanon}
V(\vp)= { V_0\over 1+\left ({ \mu^2\over \vp^2 }\right)^{k\over 2} }  = { V_0\over 1+\left ({ \mu^2\over \vp^2 }\right)^{1\over n- 2} }\ .
\ee
This is a typical KKLTI potential. 
A supergravity Jordan frame version of this new $\xi$-attractor model is based on 
\be
K(T, \overline{T}) =   {\mu^2  \over  2 ( {T} \overline{T} )^{n-2\over 2}}, \quad \Omega(T)= 1+\xi (T\overline{T})^{p\over 2}, \quad V_E= {V_0\over 1+ (T\overline{T})^{1\over 2}} \ .
\ee
Note that whereas the full Lagrangian of this theory in the Jordan frame \rf{JordanPolcan} depends on the specific choice of $\Omega(T)= 1+\xi (T\overline{T})^{p\over 2}$, the cosmological predictions of this class of models depend only on the choice of $K(T, \overline{T})$ (or $K_E (T)$) and $V_{E}$.

One may also consider the class of models with $0<n< 2$, and  change the potential
\be
V_{E} = {V_{0}\over 1+T^{-n}} \ .
\label{n1pot}\ee
The corresponding $V(\vp)$ potential is also  given by equation \rf{KKLTcanon}, but now 
\be
k = {2n\over 2-n}, \quad n= {2k\over k+2},\quad \mu^2 =  {4\over (n-2)^2\xi}\ .
\ee
For $0<n< 2$, the parameter $k$ also  scans the full range $0< k< \infty$.

As discussed in Section \ref{Sec:A}, these potentials are well defined for all $\vp$ if $k > 1$, but for $k < 1$ the derivatives of these potentials diverge at the minimum of the potential at $\vp = 0$. To avoid this problem, one can generalize these potentials to the potentials \rf{kk}, which are regular at $\vp = 0$ for all $k > 0$.  
Alternatively, one may consider a simple potential 
\be
V_{E} = V_{0 }\left(1-{T\over 2}  \right)^{2}  = \left(1- {1\over 2} \left ({ \mu\over \vp }\right)^{k}\right)^{2} \ ,
\label{kx}\ee
where $\mu = {2\over \sqrt \xi} $. This potential is also well defined for all $k > 0$.  It coincides with the potential  \rf{kkk} shown in Fig. \ref{VarMU2}.  It can be implemented in supergravity with 
\be
V_{E} =V_{0 } \left(1-\tfrac{1}{2 } (T\overline{T})^{1\over 2}   \right)^{2} \ .
\ee

It is instructive to compare the potential \rf{kx} for $n = 1$, $k = 2$ with the potential \rf{largephi} in the $\xi$-attractor model \rf{simple} introduced in  \cite{Kallosh:2025rni}.  For $\xi = 1$ the potential \rf{kx} is given by
\be\label{xixi}
V_{E} = V_{0 } \left(1- {4\over \vp^2}\right)^{2}  =  V_{0 } \left(1- {8\over \vp^2}+ O(\vp^{-4}) \right)^{2} \ .
\ee
At $\vp \gg 1$, this result coincides with \rf{largephi}. However, in the model introduced in \cite{Kallosh:2025rni}, this result holds only for $\xi = O(1)$; in the large-$\xi$ limit, the model makes the same prediction as the Starobinsky model. Meanwhile, the expression \rf{kx} remains valid for all values of $\xi$, and for all values of $k >0$.  In the large $\xi$ limit, it predicts $n_{s} = 1- 3/2 N \sim 0.973$ for $N = 55$.  By decreasing $k$, one can increase $n_{s}$ all the way up to $1-1/N$, i.e., up to 0.980 for $N = 50$ or   0.982 for $N = 55$.

\section{Discrete B-mode  targets in exponential and polynomial $\xi={1\over 6\a}$ attractors}\label{Sec:Range}

In our general definition of exponential and polynomial attractors in equations  \rf{exp} and \rf{polynomial}, we have a single parameter $\a$ in the exponential case and two parameters $(\a, k)$ or $(\mu,k)$ in the polynomial case.  

Regarding future B-mode experiments, we would like to specify the values of $\a$ and $\mu$ for various models, as these parameters determine $r$ via the equations \rf{a}, \rf{KKLTIw}, and their $\xi$ counterparts.\footnote{Our new $(\xi, n)$ attractors have some common features with Palatini attractors \cite{Jarv:2017azx,Tenkanen:2020dge,Barman:2023opy}. Specifically, in Palatini attractors the term $-{3\over 2} {(\Omega')^2\over \Omega}$ is also absent in the expression for $K_{E}$, and $r\to 0$ when $\xi\to \infty$. However, our models are based on standard general relativity with a metric formalism and can be generalized to supergravity, unlike models with independent affine connections; see \cite{Kallosh:2026tjm} for a more detailed discussion.}

 Hyperbolic $n=2$ case Sec.  \ref{Sec:A}:Kallosh:2026tjm
\bea
&&r\sim \a={1\over 6\xi}  \qquad \hskip 1.6 cm {\rm exponential} \, \, {\rm attractors}\cr
&& r\sim \a^{k\over k+2}=  \Big ({1\over 6\xi}\Big )^{k\over k+2}  \qquad {\rm polynomial} \, \,{\rm attractors}
\nonumber \eea

 Polynomial  $n\neq 2$ case Sec.  \ref{Sec:B}:
\bea
\hskip 1 cm   r\sim \mu^{2k\over k+2} \sim \left({1\over \xi} \right)^{k\over k+2} \hskip 1.4cm {\rm general \, \, polynomial \, \, attractors}
\nonumber 
\eea
We will now discuss only $n=2$ exponential and polynomial $\a$ attractors, where $\mu^2={3\a\over 2}$, presented in Sec. \ref{Sec:A}.

In supergravity-associated models of exponential and polynomial attractors, we study   models with  geometric kinetic terms where ${3\a \over 2}= {1\over |\mathcal{R}|}$,  and 
$\mathcal{R}$ is the \K  curvature of the underlying  hyperbolic geometry.
The reason for this choice is the  original supergravity T-model  \cite{Kallosh:2013yoa} with disk variables with the kinetic term $-3\a\ {dZ d\bar Z\over (1-Z\bar Z)^2} $ and \K potential $-3\a \ln(1-Z\bar Z)$, or an E-model with half-plane variables \cite{Cecotti:2014ipa} with the kinetic term $-3\a\ {d T d\overline{T}\over (T+\overline{T})^2} $ and \K potential $-3\a \ln(T+\overline{T})$. In both cases, there is an underlying hyperbolic geometry, but in different coordinates of the moduli space;  in both cases, the pole order is exactly $2$. In supergravity, in both cases, the scalar is a complex field; however, during inflation, only a single scalar, the inflaton, evolves. Its partner, the axion in these models, is stabilized during inflation.

\begin{figure}[H]
\centering
\includegraphics[scale=0.34]{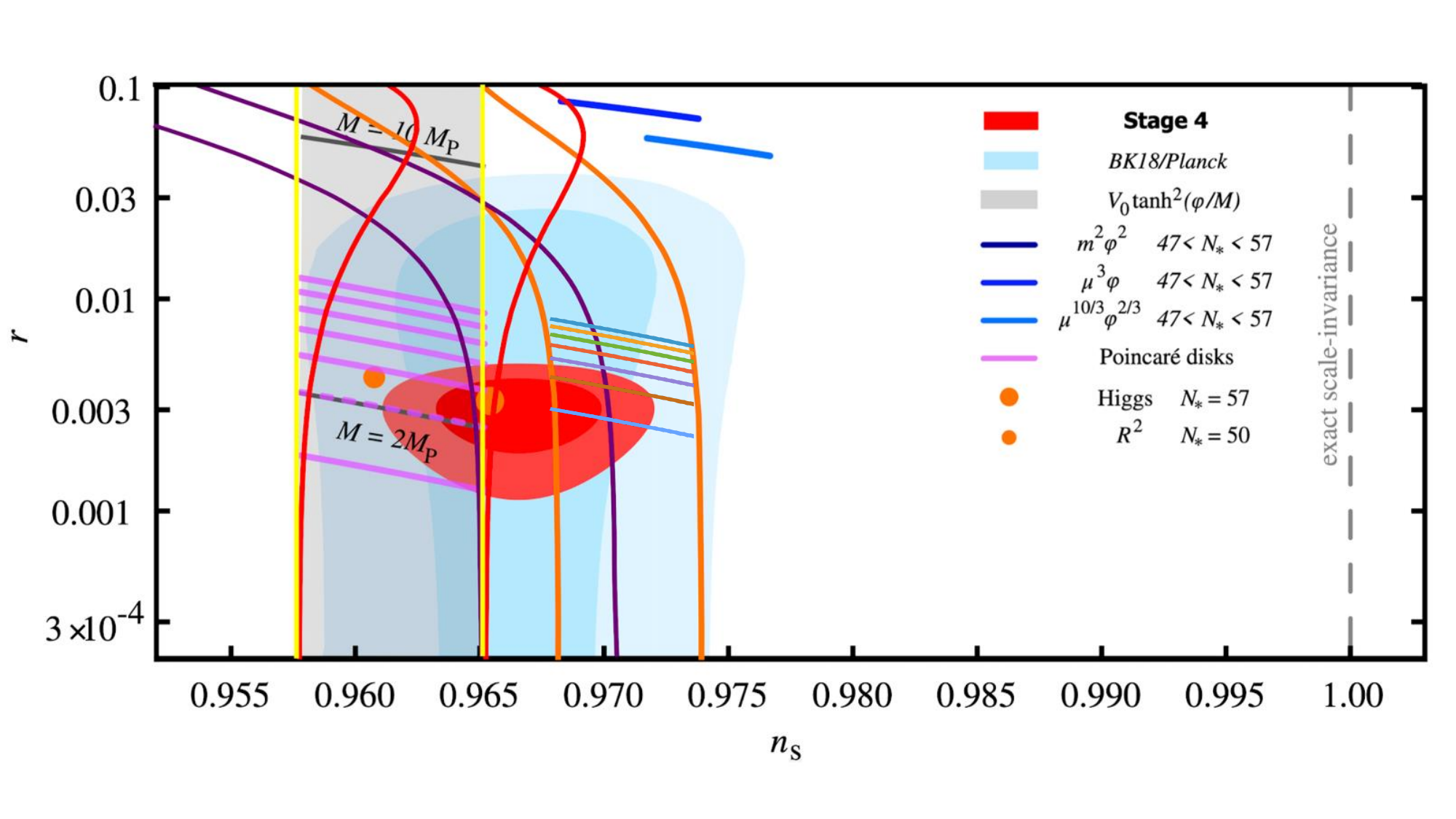}
\vskip -5pt
\caption{\footnotesize This is an extended version of figure 2  in   \cite{Chang:2022tzj} and figure 1 in \cite{Kallosh:2022vha}. It shows the predictions of exponential T-model $\alpha$ or $\xi=1/6\a$ attractors with unconstrained values of $\alpha=1/6\xi$ (gray area), E-models (red lines), predictions for $ 3\alpha={1\over 2\xi} =7,6,5,4,3,2,1$ (purple lines), as well as Higgs inflation and $R^2$ inflation (orange dots). In addition, two dark purple lines show the predictions of the polynomial $\alpha=1/6\xi$ attractors for $k = 4$, and the orange lines correspond to $k = 2$. For the last case, we also plotted the series of discrete predictions for $3\a={1\over 2\xi}=7,6,5,4,3,2,1$. The predictions are for $47 < N< 57$.}
\label{ladder}
\end{figure}

The $\xi$ counterparts of these hyperbolic models are defined in eq. \rf{hyperE} here and replace $3\a$ by a ${1\over 2\xi}$.
In these supergravity models, the \K  curvature 
\be
\mathcal{R}_{K}=- {2\over 3\a}= -4\xi
\ee 
is not restricted and therefore we see in T- and E-models in Fig. \ref{ladder} that $r$ goes down continuously all the way to $r \to 0$, $\a\to 0$, $\xi\to \infty$, $|\mathcal{R}|\to \infty$. Also in polynomial  hyperbolic attractors where  $\mu^2={3\a\over 2}= {1\over 4 \xi}$
 in \cite{Kallosh:2022feu} there is no restriction on $\mu$, it can go down to $\mu^2={3\a\over 2}= {1\over 4 \xi}\to 0$. These are examples in Fig. \ref{ladder} with $k=4,2$.

 The current bound on $r$ places an upper bound on the parameters in these models. The latest bounds on $r$ are given in \cite{Balkenhol:2025wms}.  The  constraints from the SPA+BK combination are
$
r < 0.034$, 
$n_s = 0.9682 \pm 0.0032
$.
These are important for exponential attractors (the lhs of Fig. \ref{ladder}). 
For the SPA+BK+DESI combination the constraints  in \cite{Balkenhol:2025wms} are 
$
r < 0.035$, 
$n_s = 0.9728 \pm 0.0029$.
These are important for polynomial attractors (the rhs of Fig. \ref{ladder}).

In both exponential and polynomial attractors in supergravity with hyperbolic geometry and $\mu^2={3\a\over 2}= {1\over 4 \xi}$, there is a special class of models associated with higher-dimensional 11D or 10D supergravity compactified to 4 and string theory \cite{Ferrara:2016fwe,Kallosh:2017ced}. In these models $3\a={1\over 2\xi}$ is an integer 
\be
3\a={1\over 2\xi}= 7,6,5,4,3,2,1
\ee

 We discussed this feature in \cite{Kallosh:2025ijd}; here we present the discrete targets for hyperbolic polynomial $k=2$ attractors in Fig. \ref{ladder}. It is interesting that, in upper cases, with  $3\a={1\over 2\xi} = 7$, the discrete targets for exponential and polynomial attractors are at about the same level. For smaller $\alpha$, e.g. for $3\a= {1\over 2\xi} = 1$, the difference is significant. This can be explained by the difference in the universal formulas in the exponential and polynomial cases
\be
r_{exp}= 4 {3\a\over N^2} \ , \qquad r_{pol; \,k=2}= {\sqrt{3\a}\over N^{3/2}} \ .
\ee

\begin{figure}[H]
 \centering
\includegraphics[scale=0.42]{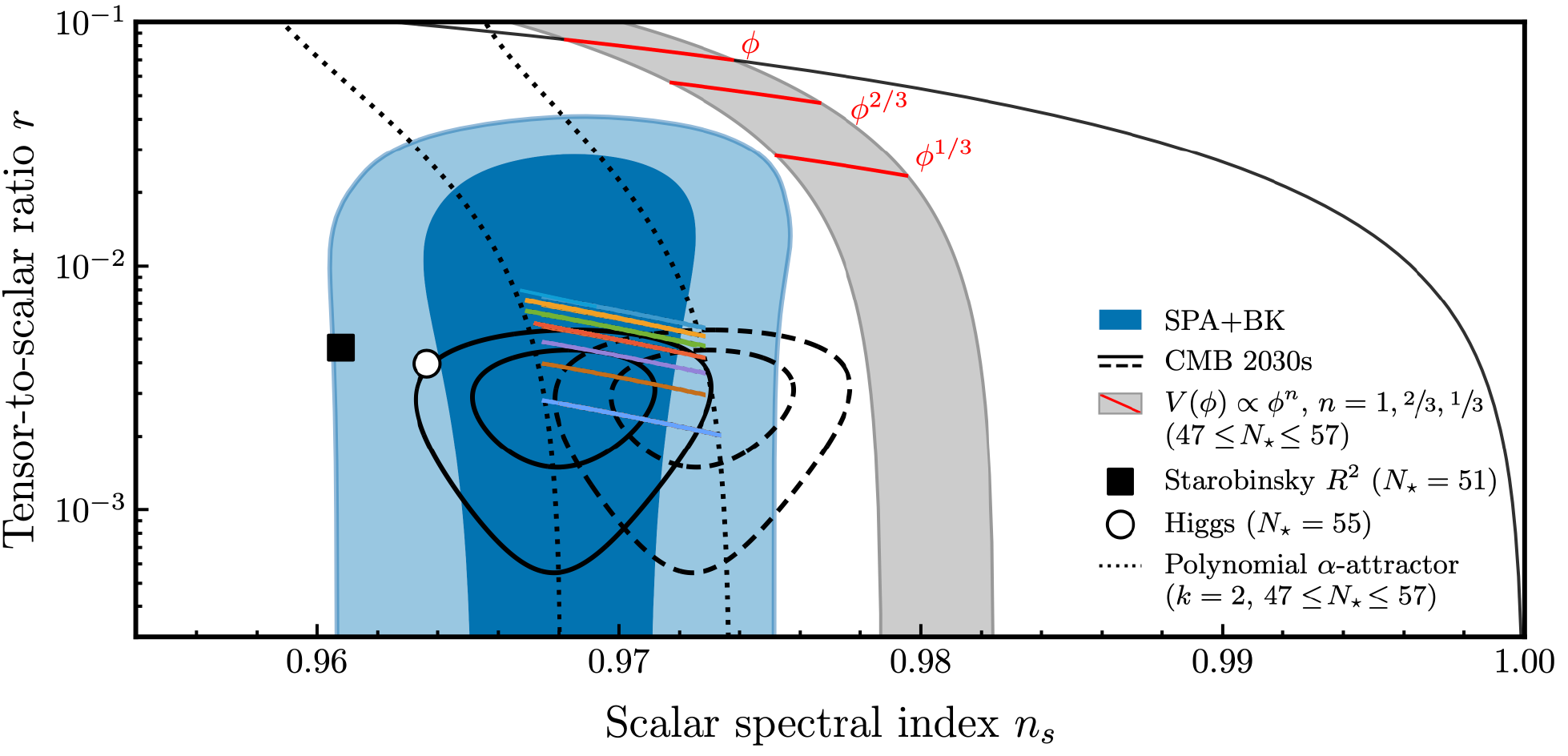}
\vskip -5pt
\caption{\footnotesize  This is an $r, n_s$ plot in  \cite{Balkenhol:2025wms}  Predictions of our polynomial $\a$ attractor model  at $k = 2$, $N_*=57$ is $n_s\approx 0.9736$  are shown as a dotted black lines.
 We also added the 7 discrete targets for the polynomial $\alpha$-attractors with $3\alpha = 7,6,5,4,3,2,1$. }
\label{k98}
\end{figure}

In Fig. \ref{k98} we present the 2025 plot of $r, n_s$  in  \cite{Balkenhol:2025wms} based on  Planck, ACT, SPT, and BICEP/Keck. It shows two forecasts for CMB constraints that may be achieved in the next decade with $10^3 r=3\pm1$ at different values of $n_s$. The one in black solid contours is based on a central value of Planck, SPT, ACT,  BICEP/Keck, and the one with black dashed contours is based on a central value of Planck, SPT, ACT,  BICEP/Keck, and DESI. One can see that the simple case of polynomial $\a$ attractors with $k=2$ and $47 \leq N \leq 57$ fits both cases.

\section{Summary  }

In this paper, we have introduced a large class of new $(\xi,n)$-attractors in the standard general relativity context with scalar fields non-minimally coupled to gravity and with the kinetic terms   $K_{E} \sim {1\over \xi T^{n}}$. These models are equivalent to exponential and polynomial $\a$ attractors, or to general polynomial $(\mu, k)$ attractors in models with minimal coupling to gravity.

The relation between models with non-minimal coupling to gravity in the Jordan frame and the Einstein frame with minimal coupling of scalars to gravity is based on a superconformal approach to supergravity developed recently with regard to new $\xi$-attractors in \cite{Kallosh:2026tjm}. In this approach, the Einstein and Jordan frames are two distinct gauge-fixing choices for the Weyl and other superconformal symmetries.

For the models with a single  complex scalar, the bosonic part of the Einstein frame supergravity Lagrangian is
\begin{eqnarray}
{\mathcal{L}_{E}(T, \overline{T})\over \sqrt {-g_E}} = \frac{1}{2} {R}\left (g_E\right)
-g_{T \overline{T}}  {\partial}_\mu T  {\partial}^\mu \overline{T}
-V_E (T, \overline{T}) \ ,
  \label{Turin-41E}
\end{eqnarray}
where the metric is a second derivative of the \K potential, $g_{T \overline{T}}= \partial_T \partial _{\overline{T}} K(T, \overline{T})$.  This defines $K_E$ in eq. \rf{Einstein}
\be\label{KKKK}
{1\over 2} K_E(T,\bar T)= g_{T \overline{T}}|_{_{T=\overline{T}}} \ .
\ee

The Jordan frame action, once the choice of the frame function $\Omega (T, \overline{T})$ is made, has the universal form which we will present here at $T=\overline{T}$,  with $\Omega (T, \overline{T}) = \Omega(T)$: 
\begin{eqnarray}
{\mathcal{L}_{J}(T)\over \sqrt {-g_J}}=  \frac{\Omega}{2} {R}\left (g_J\right)  -{1\over 2}K_J(T)  ({\partial} T )^2 - \frac{\Omega^2}{V_E (T) }\ ,
  \label{Turin-41}
\end{eqnarray}
where 
\be
{1\over 2}K_J(T)= \left (\Omega \, g_{T \overline{T}}  -3 {\partial_T \Omega  \, \partial_{\overline{T}} \Omega \over \Omega }\right )|_{_{T=\overline{T}}} = {1\over 2} \Big (\Omega K_E - {3\over 2} {(\Omega')^2\over \Omega} \Big)\ .
\label{KJs}\ee

Thus, if we start with the Einstein frame and know the \K potential $K(T, \overline{T})$ or the \K metric $g_{T \overline{T}}$ at $T=\overline{T}$,  we can find $K_J(T)$ according to eq. \rf{KJs}. And we have a Jordan frame action \rf{Turin-41}, for any choice of the function $\Omega$, which is guaranteed to be (classically) equivalent to the Einstein frame theory in eq. \rf{Turin-41}. For example, we can make a choice of the frame function $\Omega(T)=1+\xi T^n$ so that the Jordan frame action \rf{KJs} depends only on $(\xi, n)$. However, the Einstein frame action \rf{Turin-41E} does not depend on the choice of $\Omega(T)$ in the Jordan frame action \rf{KJs}. The cosmological predictions of these models depend only on the choice of $K_E$ and $V_{E}$, and only on two parameters $\xi$ and $n$, independently of any specific choice of $\Omega$ in \rf{Turin-41}.  

In this paper, we considered the simplest models of that type, with \K potentials $K(T, \overline{T}) \sim \ln (T+\overline{T}) $ and ${\mu^2  \over  2 ( {T} \overline{T} )^{1\over k}}$. These models can be easily generalized using the streamlined approach to supergravity, which allows one to construct a theory with an arbitrary \K potential $K(T, \overline{T})$ and an arbitrary potential $V(T,\overline{T})$ \cite{Kallosh:2025dac}.  Some particular examples considered in  \cite{Kallosh:2025dac} involved \K potentials $\sim \ln (1- Z\bar Z)$ and $(1-Z\bar Z)^{2-n}$.

Using this kind of supergravity guidance, we have found a large class of models with non-linear coupling to gravity that are equivalent to some well-known inflationary attractor models in theories with minimal coupling to gravity.

In the case of hyperbolic geometry ($n = 2$), we found that the new exponential and polynomial $\xi$-attractors  are equivalent to corresponding $\a$-attractors \cite{Kallosh:2013yoa,Kallosh:2022feu} with  a  relation between these two cases
\be
\alpha= {1\over 6\xi}\, .
\ee
 In particular, this means that in new $\xi$-attractors  the strong coupling limit $\xi \to \infty$ corresponds to the limit $ \a\to 0$.
This new relation is different from 
 earlier models of $\xi$-attractors in \cite{Galante:2014ifa}, where this relation was  $\a= 1+ {1\over 6\xi}$ and in particular,   $\xi \to \infty \, \,  \Rightarrow \, \,   \a\to 1$.

We have presented the KKLTI polynomial attractor potentials  \rf{KKLTIk}, which are well-defined for all $k >1$, see Figs. \ref{KKLTIpot} and \ref{VarMU}.  We also presented the generalized potentials \rf{kk} and \rf{kkk}, which are valid for any $k > 0$. The potentials \rf{kk} are similar to the KKLTI potentials  \rf{KKLTIk}, but they are well-defined at $\vp = 0$ even if $k < 1$. The potentials \rf{kkk} are shown in Fig. \ref{VarMU2}. These new potentials can describe models with $n_{s}$ all the way up to $1-1/N \sim 0.98$ for $N \sim 50$. We have provided supergravity versions for all new $\xi$-attractors proposed in this article.

The future detection of the B-models in the context of cosmological $\a={1\over 6\xi}$ attractors will be the measurement of the \K curvature $\mathcal{R}_{K}=- {2\over 3\a}= -4\xi$ of the underlying moduli space geometry for exponential and polynomial attractors, respectively.
\be
{\rm exponential:}   \qquad r \sim {1\over |\mathcal{R}_{K}|} \qquad \qquad {\rm polynomial:}  \qquad r \sim {1\over |\mathcal{R}_{K}|^{k\over k+1}}
\ee
The higher the curvature $\mathcal{R}_{K}$, the smaller the level of gravitational waves $r$ in these models.

 In Figs. In \ref{ladder} and \ref{k98}, we have shown the discrete targets for B-mode detection for both exponential and polynomial $\a={1\over 6\xi}$ attractors. The ones for the exponential attractors in  Fig. \ref{ladder} are known as Poincar\'e disks targets for the LiteBIRD \cite{LiteBIRD:2022cnt}. The polynomial attractors originating from the models with $n = 2$, $\a={1\over 6\xi}$ also have discrete targets, for any $k$. Seven such targets are shown in Fig. \ref{ladder}  between the two orange lines,  and in Fig. \ref{k98} for $k=2$ polynomial $\a$ attractor.  
 
 Note that the upper discrete target for exponential alpha attractors with $\alpha = 7/3$ has $r = {7\over 3} \ {12\over N^{2}}$, which is $7/3$ times higher than the predictions of the Starobinsky model and Higgs inflation $r = {12\over N^{2}}$. A similar statement is true for the upper discrete target for $k = 2$, $\alpha = 7/3$.  
 We conclude that the set of cosmological attractors discussed in this paper can match any combination of Planck, BICEP/Keck, ACT, SPT, and DESI  data, and simultaneously provide us with a set of new discrete targets that can be tested by the B-modes search much earlier than the more familiar targets with $\alpha = 1$.

\

{\bf Acknowledgments:} We are grateful to 
Raphael Flauger,  Marek Olechowski, Stefan Pokorski,  Diederik Roest, and Yusuke Yamada for insightful discussions, and to Tommi Tenkanen and Javier Rubio for informing us about Palatini attractors. This work is supported by Leinweber Institute for Theoretical Physics at Stanford and by NSF Grant PHY-2310429.

\bibliographystyle{JHEP}
\bibliography{lindekalloshrefs}
\end{document}